\documentclass[aps,prd,twocolumn,superscriptaddress,showpacs,preprintnumbers,amsmath,amssymb,floatfix]{revtex4-1}

\usepackage{subfigure}
\usepackage{graphicx} 
\usepackage{dcolumn}  

\graphicspath{{ps}}

\begin{document}



\title{ \quad\\[1.0cm]Measurement of Branching Fractions of Hadronic Decays of the $\Omega_c^0$ Baryon}

\noaffiliation
\affiliation{University of the Basque Country UPV/EHU, 48080 Bilbao}
\affiliation{Beihang University, Beijing 100191}
\affiliation{Budker Institute of Nuclear Physics SB RAS, Novosibirsk 630090}
\affiliation{Faculty of Mathematics and Physics, Charles University, 121 16 Prague}
\affiliation{University of Cincinnati, Cincinnati, Ohio 45221}
\affiliation{Deutsches Elektronen--Synchrotron, 22607 Hamburg}
\affiliation{University of Florida, Gainesville, Florida 32611}
\affiliation{Department of Physics, Fu Jen Catholic University, Taipei 24205}
\affiliation{Justus-Liebig-Universit\"at Gie\ss{}en, 35392 Gie\ss{}en}
\affiliation{Gifu University, Gifu 501-1193}
\affiliation{SOKENDAI (The Graduate University for Advanced Studies), Hayama 240-0193}
\affiliation{Gyeongsang National University, Chinju 660-701}
\affiliation{Hanyang University, Seoul 133-791}
\affiliation{University of Hawaii, Honolulu, Hawaii 96822}
\affiliation{High Energy Accelerator Research Organization (KEK), Tsukuba 305-0801}
\affiliation{J-PARC Branch, KEK Theory Center, High Energy Accelerator Research Organization (KEK), Tsukuba 305-0801}
\affiliation{IKERBASQUE, Basque Foundation for Science, 48013 Bilbao}
\affiliation{Indian Institute of Science Education and Research Mohali, SAS Nagar, 140306}
\affiliation{Indian Institute of Technology Bhubaneswar, Satya Nagar 751007}
\affiliation{Indian Institute of Technology Guwahati, Assam 781039}
\affiliation{Indian Institute of Technology Hyderabad, Telangana 502285}
\affiliation{Indian Institute of Technology Madras, Chennai 600036}
\affiliation{Indiana University, Bloomington, Indiana 47408}
\affiliation{Institute of High Energy Physics, Chinese Academy of Sciences, Beijing 100049}
\affiliation{Institute of High Energy Physics, Vienna 1050}
\affiliation{Institute for High Energy Physics, Protvino 142281}
\affiliation{University of Mississippi, University, Mississippi 38677}
\affiliation{INFN - Sezione di Napoli, 80126 Napoli}
\affiliation{INFN - Sezione di Torino, 10125 Torino}
\affiliation{Advanced Science Research Center, Japan Atomic Energy Agency, Naka 319-1195}
\affiliation{J. Stefan Institute, 1000 Ljubljana}
\affiliation{Kanagawa University, Yokohama 221-8686}
\affiliation{Institut f\"ur Experimentelle Kernphysik, Karlsruher Institut f\"ur Technologie, 76131 Karlsruhe}
\affiliation{Kennesaw State University, Kennesaw, Georgia 30144}
\affiliation{Department of Physics, Faculty of Science, King Abdulaziz University, Jeddah 21589}
\affiliation{Korea Institute of Science and Technology Information, Daejeon 305-806}
\affiliation{Korea University, Seoul 136-713}
\affiliation{Kyoto University, Kyoto 606-8502}
\affiliation{Kyungpook National University, Daegu 702-701}
\affiliation{\'Ecole Polytechnique F\'ed\'erale de Lausanne (EPFL), Lausanne 1015}
\affiliation{P.N. Lebedev Physical Institute of the Russian Academy of Sciences, Moscow 119991}
\affiliation{Faculty of Mathematics and Physics, University of Ljubljana, 1000 Ljubljana}
\affiliation{Ludwig Maximilians University, 80539 Munich}
\affiliation{Luther College, Decorah, Iowa 52101}
\affiliation{Malaviya National Institute of Technology Jaipur, Jaipur 302017}
\affiliation{University of Maribor, 2000 Maribor}
\affiliation{Max-Planck-Institut f\"ur Physik, 80805 M\"unchen}
\affiliation{School of Physics, University of Melbourne, Victoria 3010}
\affiliation{University of Miyazaki, Miyazaki 889-2192}
\affiliation{Moscow Physical Engineering Institute, Moscow 115409}
\affiliation{Moscow Institute of Physics and Technology, Moscow Region 141700}
\affiliation{Graduate School of Science, Nagoya University, Nagoya 464-8602}
\affiliation{Kobayashi-Maskawa Institute, Nagoya University, Nagoya 464-8602}
\affiliation{Nara Women's University, Nara 630-8506}
\affiliation{National Central University, Chung-li 32054}
\affiliation{National United University, Miao Li 36003}
\affiliation{Department of Physics, National Taiwan University, Taipei 10617}
\affiliation{H. Niewodniczanski Institute of Nuclear Physics, Krakow 31-342}
\affiliation{Niigata University, Niigata 950-2181}
\affiliation{Novosibirsk State University, Novosibirsk 630090}
\affiliation{Osaka City University, Osaka 558-8585}
\affiliation{Pacific Northwest National Laboratory, Richland, Washington 99352}
\affiliation{Panjab University, Chandigarh 160014}
\affiliation{University of Pittsburgh, Pittsburgh, Pennsylvania 15260}
\affiliation{Theoretical Research Division, Nishina Center, RIKEN, Saitama 351-0198}
\affiliation{University of Science and Technology of China, Hefei 230026}
\affiliation{Showa Pharmaceutical University, Tokyo 194-8543}
\affiliation{Soongsil University, Seoul 156-743}
\affiliation{University of South Carolina, Columbia, South Carolina 29208}
\affiliation{Stefan Meyer Institute for Subatomic Physics, Vienna 1090}
\affiliation{Sungkyunkwan University, Suwon 440-746}
\affiliation{School of Physics, University of Sydney, New South Wales 2006}
\affiliation{Department of Physics, Faculty of Science, University of Tabuk, Tabuk 71451}
\affiliation{Tata Institute of Fundamental Research, Mumbai 400005}
\affiliation{Department of Physics, Technische Universit\"at M\"unchen, 85748 Garching}
\affiliation{Toho University, Funabashi 274-8510}
\affiliation{Department of Physics, Tohoku University, Sendai 980-8578}
\affiliation{Earthquake Research Institute, University of Tokyo, Tokyo 113-0032}
\affiliation{Department of Physics, University of Tokyo, Tokyo 113-0033}
\affiliation{Tokyo Institute of Technology, Tokyo 152-8550}
\affiliation{Tokyo Metropolitan University, Tokyo 192-0397}
\affiliation{University of Torino, 10124 Torino}
\affiliation{Virginia Polytechnic Institute and State University, Blacksburg, Virginia 24061}
\affiliation{Wayne State University, Detroit, Michigan 48202}
\affiliation{Yonsei University, Seoul 120-749}
  \author{J.~Yelton}\affiliation{University of Florida, Gainesville, Florida 32611} 
  \author{I.~Adachi}\affiliation{High Energy Accelerator Research Organization (KEK), Tsukuba 305-0801}\affiliation{SOKENDAI (The Graduate University for Advanced Studies), Hayama 240-0193} 
  \author{H.~Aihara}\affiliation{Department of Physics, University of Tokyo, Tokyo 113-0033} 
  \author{S.~Al~Said}\affiliation{Department of Physics, Faculty of Science, University of Tabuk, Tabuk 71451}\affiliation{Department of Physics, Faculty of Science, King Abdulaziz University, Jeddah 21589} 
  \author{D.~M.~Asner}\affiliation{Pacific Northwest National Laboratory, Richland, Washington 99352} 
  \author{H.~Atmacan}\affiliation{University of South Carolina, Columbia, South Carolina 29208} 
  \author{V.~Aulchenko}\affiliation{Budker Institute of Nuclear Physics SB RAS, Novosibirsk 630090}\affiliation{Novosibirsk State University, Novosibirsk 630090} 
  \author{T.~Aushev}\affiliation{Moscow Institute of Physics and Technology, Moscow Region 141700} 
  \author{R.~Ayad}\affiliation{Department of Physics, Faculty of Science, University of Tabuk, Tabuk 71451} 
  \author{T.~Aziz}\affiliation{Tata Institute of Fundamental Research, Mumbai 400005} 
% ...
\collaboration{The Belle Collaboration}


\begin{abstract}
Using a data sample of 980 ${\rm fb}^{-1}$ of $e^+e^-$ annihilation data taken with the Belle detector
operating at the KEKB asymmetric-energy $e^+e^-$ collider, we report the results of a study of the decays of the 
$\Omega_c^0$ charmed baryon into hadronic final states.
We report the most precise measurements to date of the relative branching fractions 
of the $\Omega_c^0$ into $\Omega^-\pi^+\pi^0$, $\Omega^-\pi^+\pi^-\pi^+$,
$\Xi^-K^-\pi^+\pi^+$, and
$\Xi^0K^-\pi^+$, 
as well as the first measurements of the branching fractions of the $\Omega_c^0$ into
$\Xi^-\bar{K}^0\pi^+$,
$\Xi^0\bar{K}^0$,
and
$\Lambda \bar{K}^0\bar{K}^0$, 
all with respect to the $\Omega^-\pi^+$ decay. In addition, we 
investigate the resonant substructure of these modes. Finally, we present
a limit on the branching fraction for the decay $\Omega_c^0\to\Sigma^+K^-K^-\pi^+$.
\end{abstract}


\maketitle


{\renewcommand{\thefootnote}{\fnsymbol{footnote}}}
\setcounter{footnote}{0}

\section*{Introduction}
The $\Omega_c^0$ comprises the combination of a charm quark and two strange quarks~\cite{CC}.
The ground-state $\Omega_c^0$ has the $ss$ diquark in a $J^P=1^+$ configuration, and decays weakly.
There are no measurements of the absolute branching fractions of the $\Omega_c^0$, but some measurements of
the branching ratios of modes with respect to the normalizing 
mode $\Omega^-\pi^+$ have been made~\cite{CLEO,CLEO2,BaBar}. However, because the 
production cross section of the $\Omega_c^0$ is lower than the other singly charmed baryons, and 
because it typically decays to more complicated final states, there is less information
on its hadronic decays than there is for the other weakly decaying charmed baryons ($\Lambda_c^+$, $\Xi_c^0$, and $\Xi_c^+$)
or for the charmed mesons. 

In this paper, we present the most precise measurements of the branching fractions 
of $\Omega_c^0$ decays into the four decay modes ($\Omega^-\pi^+\pi^0$, $\Omega^-\pi^+\pi^-\pi^+$,
$\Xi^-K^-\pi^+\pi^+$,
$\Xi^-\bar{K}^0\pi^+$). These modes have previously been measured by the 
CLEO~\cite{CLEO} and/or BaBar~\cite{BaBar} Collaborations. We also present the measurement of 
three previously unreported
decays ($\Xi^-\bar{K}^0\pi^+$, $\Xi^0\bar{K}^0$ and $\Lambda\bar{K}^0\bar{K}^0$) 
and a search for one other decay, $\Sigma^+K^-K^-\pi^+$, 
that was reported by the E687 Collaboration~\cite{E687}. 
All branching fractions are measured relative to the decay $\Omega_c^0 \to \Omega^-\pi^+$.
In addition, we investigate the resonant substructure of the decays 
we observe. The choice of decay modes was guided by previous observations, analogy with other charmed baryon decay
modes, and consideration of the detector capabilities.

The four ground-state charmed baryons all decay predominantly through the weak decay $c\to sW^+$, 
but each has its own features.
Uniquely among the four, the two spectator quarks of the $\Omega_c^0$ have the same flavor, and this leads to many
decay diagrams producing the same final states. Constructive interference among these diagrams 
is thought to explain the short lifetime, 
despite the fact that, 
unlike the $\Lambda_c^+$ and $\Xi_c^0$, the $\Omega_c^0$ cannot decay via a Cabibbo-favored W-exchange diagram~\cite{GUBERINA}.
Measuring the branching fractions of all the charmed hadrons helps disentangle the various processes involved and
adds to our knowledge of the dynamics of charmed baryon decays.

This analysis uses a data sample of $e^+e^-$ annihilations recorded by the Belle detector~\cite{Belle} 
operating at the KEKB asymmetric-energy $e^+e^-$
collider~\cite{KEKB}. It corresponds to an integrated luminosity of 980 ${\rm fb}^{-1}$.
The majority of these data were taken with the accelerator energy tuned for production of the $\Upsilon(4S)$ resonance, as this is optimum
for investigation of $B$ decays.
However, the $\Omega_c^0$ particles in this analysis are produced in continuum charm production and are of
higher momentum than those that are
decay products of $B$ mesons, so
the dataset used in this analysis also includes the Belle data taken at beam energies corresponding to the other $\Upsilon$ resonances
and the nearby continuum ($e^+e^- \to q\bar{q}$, where $q \in \{u,\ d,\ s,\ c\}$).

\section*{The Belle Detector and Particle Reconstruction}

The Belle detector is a large-solid-angle spectrometer comprising six sub-detectors: the Silicon Vertex Detector (SVD), the 50-layer Central
Drift Chamber (CDC), the Aerogel Cherenkov Counter (ACC), the Time-of-Flight scintillation counter (TOF), 
the electromagnetic calorimeter, and the 
$K_L$ and muon detector. A superconducting solenoid produces a 1.5 T magnetic field throughout the first five of these sub-detectors.
The detector is described in detail elsewhere~\cite{Belle}. Two inner detector configurations were used. 
The first comprised a 2.0 cm radius beampipe and a 3-layer silicon vertex detector, and the second a 1.5 cm radius beampipe and a 4-layer silicon detector and a small-cell inner drift chamber.

Final-state charged particles, 
$\pi^{\pm}, K^{-}$, and $p$, are selected using the likelihood 
information from the tracking (SVD, CDC) and charged-hadron identification (CDC, ACC, TOF) systems, 
${\cal L}(h1:h2) = {\cal L}_{h1}/({\cal L}_{h1} + {\cal L}_{h2})$, 
where $h_1$ and $h_2$ are $p$, $K$, and $\pi$ as appropriate. In general, we require proton candidates 
to have ${\cal L}(p:K)>0.6$ and ${\cal L}(p:\pi)>0.6$ ($\approx 96\%$ efficient); 
kaon candidates to have ${\cal L}(K:p)>0.6$ and ${\cal L}(K:\pi)>0.6$ ($\approx 94\%$ efficient); 
and pions to have the less restrictive requirements of
${\cal L}(\pi:K)>0.2$ and ${\cal L}(\pi:p)>0.2$ ($\approx 99\%$ efficient). 
The $\pi^0$ candidates used in hyperon reconstruction are formed from two clusters unassociated 
with a charged track, each consistent with being
due to a photon, and each of energy above $50\ {\rm MeV}$ in the laboratory frame. The invariant mass of the photon pair is 
required to be
within 3 standard deviations ($\sigma$) of the $\pi^0$ mass~\cite{PDG}. Because of the large combinatorial background,
the $\pi^0$ candidates used for $\Omega_c^0 \to \Omega^-\pi^+\pi^0$ reconstruction have more restrictive requirements of 
at least 
100 ${\rm MeV}$ energy per photon, at least $300\ {\rm MeV}/c$  $\pi^0$ momentum, 
and an invariant mass within $2\sigma$ of the $\pi^0$ nominal mass.

The $\Lambda\ (K_S^0)$ candidates are reconstructed from 
$p\pi^-\ (\pi^+\pi^-)$ pairs with a production vertex significantly
separated from the
nominal interaction point (IP) in the $r-\phi$ plane (perpendicular to the beam axis). 
For the case of the proton from the $\Lambda$, the particle identification (PID) is loosened to 
${\cal L}(p:K)>0.2$ and ${\cal L}(p:\pi)>0.2$.  
The $\Lambda$ candidates used as immediate daughters of $\Xi_c$ candidates are required to have trajectories consistent with 
origination at the IP, 
but those that are daughters of $\Xi^-$, $\Xi^0$ or $\Omega^-$ candidates do not have this requirement.

The $\Xi^-$ and $\Omega^-$ candidates are reconstructed from the $\Lambda$ candidates detailed above, 
together with a $\pi^-$
or $K^-$ candidate. The vertex formed from the $\Lambda$ and $\pi/K$ is required to be at a smaller radial distance
from the IP than the $\Lambda$ 
decay vertex. 

The $\Xi^0$ and $\Sigma^+$ reconstruction is complicated by the fact that the parent hyperon decays with a $\pi^0$ (which has
negligible vertex position information) as one of its daughters. 
In the case of the $\Sigma^+ \to p \pi^0$ reconstruction, combinations
of $\pi^0$ candidates and protons are made using those protons with a
significantly large ($>1$ mm) distance of closest approach (DOCA) to the IP. Then, taking the IP as the point of 
origin of the $\Sigma^+$, the point of intersection of the $\Sigma^+$ trjectory and the reconstructed proton trajectory is found. 
This position is taken as the decay location of the $\Sigma^+$ hyperon, and the $\pi^0$ is then re-fit using this as its point of 
origin. Only those combinations with the decay location of the $\Sigma^+$ indicating a positive $\Sigma^+$ pathlength are retained.
The $\Xi^0$ is reconstructed in a similar manner, but it is not necessary to require a large
DOCA with respect to the IP.

Mass requirements are placed on all the hyperons reconstructed, based on the nominal masses of these particles~\cite{PDG}. 
The half-widths of the allowed ranges of these mass requirements, 
all corresponding to approximately two standard deviations of the resolution,  
are
8.0, 5.0, 3.5, 3.5, and 3.5 ${\rm MeV}/c^2$
for $\Sigma^+$, $\Xi^0$, $\Xi^-$, $\Omega^-$, and $\Lambda$, 
respectively. The particles 
are then kinematically constrained to the expected masses for further analysis.

\section*{${\mathbf \Omega_c^0}$ Reconstruction}

Baryons and mesons detailed above are combined to reconstruct $\Omega_c^0$ candidates. 
Once the daughter particles of a $\Omega_c$ candidate are selected, the $\Omega_c$ candidate itself is made by kinematically fitting the 
daughters to a common decay vertex. The IP is not included in this vertex, as the small decay length associated 
with the $\Omega_c$
decays, though very short compared with the $\Xi^-$, $\Xi^0$, $\Omega^-$, and $\Sigma^+$ decay lengths, is not negligible. The 
$\chi^2$ of this vertex fit is required to be consistent with all the daughters being produced by a common parent.
To reduce combinatorial background, we require a scaled momentum
of $x_p > 0.6$, where $x_p = p^*c/\sqrt{(s/4 - m^2c^2)}$, $p^*$ is the momentum of the $\Omega_c$ candidate 
in the $e^+e^-$ center-of-mass frame, $s$ is the total
center-of-mass energy squared, and $m$ is the reconstructed mass.
Charmed baryons are known to have a hard fragmentation function, and this requirement 
produces a good signal-to-noise ratio while retaining 
high signal efficiency.

\begin{figure}[htb]                                                                                                                   

\includegraphics[width=3.5in]{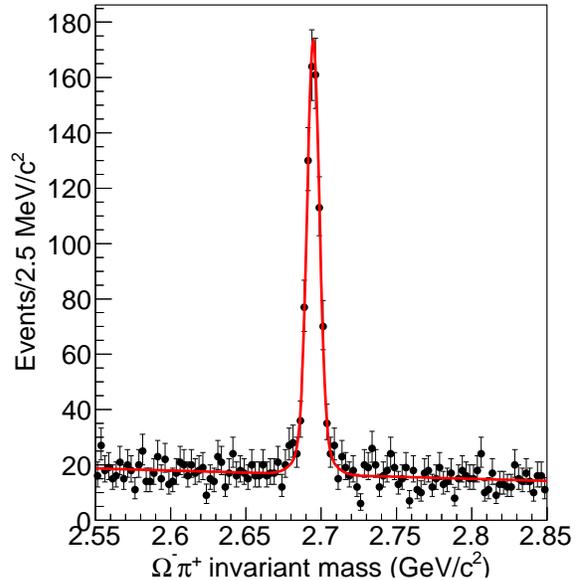}
\caption{ Invariant mass distribution for the normalizing mode $\Omega_c^0\to\Omega^-\pi^+$. 
The fit is described in the text.}
\label{fig:Figure1}

\end{figure}

Figure~\ref{fig:Figure1} shows the invariant mass distribution for the normalizing mode $\Omega_c^0 \to \Omega^-\pi^+$. 
A double-Gaussian signal function together with 
a first-order polynomial function to represent the background are fit to this distribution.
For this and all similar distributions in this analysis, 
the resolution function is obtained by studying Monte Carlo (MC) events 
generated using EvtGen~\cite{EVTGEN}, and having the Belle detector response simulated using GEANT3~\cite{GEANT3}. 
Taking the measure of each width to be the weighted average of the widths of the two Gaussian functions of the resolution function, 
the ratio of the width found by fitting the 
data in this channel to that found by fitting the MC is $1.035\pm0.045$. This confirms that the 
MC simulation predicts the resolution well.
\begin{figure*}[htb]                                                                                                                   
\includegraphics[width=7.0in]{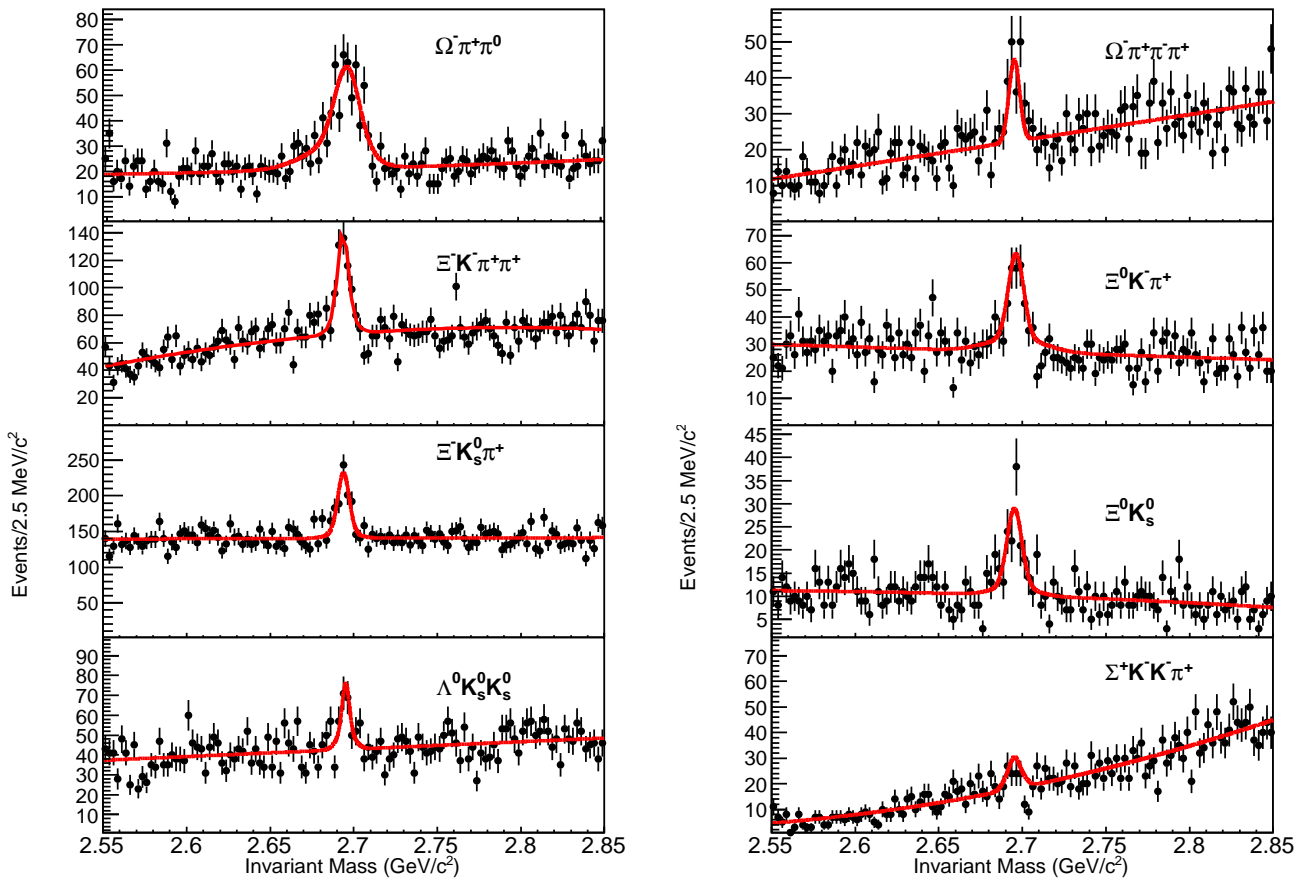}
\caption{ Invariant mass distributions for the eight modes under consideration. The fits are described in the text.}
\label{fig:Figure2}
\end{figure*}
 
Figure~\ref{fig:Figure2}
shows the invariant mass distributions for the other eight $\Omega_c^0$ decay modes under consideration. 
A fit is made to each distribution comprising the sum
of a double-Gaussian signal function, as obtained from MC, and a Chebyshev polynomial background function
whose order is the lowest that allows
a satisfactory fit. An exception is the case of the
$\Omega^-\pi^+\pi^0$ final state, for which the resolution function is a bifurcated Gaussian to account  
for the asymmetry in the mass
distribution found in MC. 
With the exception of the mode $\Omega_c^0\to\Sigma^+K^-K^-\pi^+$, 
the masses in the fits are free parameters; nevertheless, the resultant masses are consistent with the world-average~\cite{PDG}, 
which is dominated by the measurement in a previous Belle analysis using a subset of the data
presented here~\cite{Belle2}. In all cases, the resolution functions are fixed from the MC simulation, but should their 
widths be allowed
to float, each would have a width within two standard deviations of the MC values.

The yields and statistical uncertainties for each mode are listed in Table~\ref{tab:yields}, together with the resolution 
and the order of the polynomial background function used. The efficiencies, obtained from the MC simulation, include all 
branching fractions of the subsequent decays~\cite{PDG}.
In the cases where significant substructure is observed (as described in the next section), 
the MC is generated with this substructure
included. This last effect does not change the efficiency of any mode by more than 3\% of its nominal value.

\begin{table}[htb]
\caption{The summary of the results of the fits shown in Figs.~\ref{fig:Figure1} and \ref{fig:Figure2}. } 

\begin{tabular}
 {     c          |c             |c       |c      |c     }

\hline \hline
Mode                     & Signal & Order of & Resolution  & Efficiency \\
                         & yield & polynomial &(${\rm MeV}/c^2$) & (\%) \\

\hline
$\Omega^-\pi^+     $      & $691\pm29$   & 1  & 5.1 & 10.08 \\
$\Omega^-\pi^+\pi^0$      & $403\pm31$   & 2  & 13.3 & 2.95 \\
$\Omega^-\pi^+\pi^-\pi^+$ & $108\pm16$  & 1  & 4.4 &5.23 \\
$\Xi^-K^-\pi^+\pi^+$      & $278\pm27$  & 2  & 4.3 &5.98 \\
$\Xi^0K^-\pi^+$           & $168\pm21$  & 1  & 7.8 &2.09 \\
$\Xi^-{K^0_S}\pi^+$     & $349\pm36$  & 1  & 4.6  & 4.81 \\
$\Xi^0{K^0_S}$            &$98\pm15$  & 2  & 7.0  & 1.73\\
$\Lambda{K^0_S}{K^0_S}$ &$95\pm18$ & 1  & 3.7  & 3.22     \\
$\Sigma^+K^-K^-\pi^+$     & $17\pm8$  & 2  &3.8 & 2.00 \\

\hline 
\hline

\end{tabular}

\label{tab:yields}
\end{table}

\section*{Resonant Substructure}

\begin{figure*}[htb]                                                                                                                   
\includegraphics[width=6in]{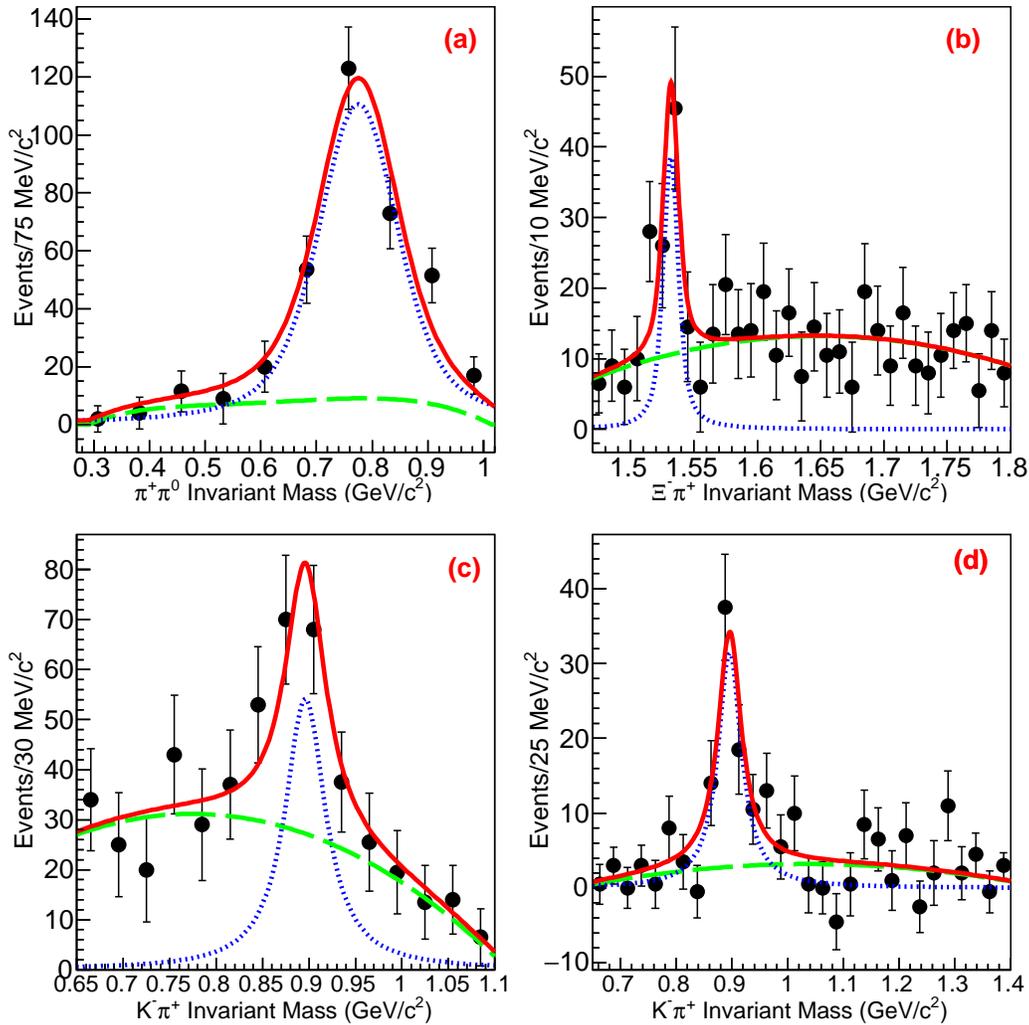}
\caption{ Background-subtracted invariant mass distributions for two particle combinations: (a) $\pi^+\pi^0$ for $\Omega_c^0\to\Omega^-\pi^+\pi^0$
decays, (b) $\Xi^-\pi^+$ and (c) $K^-\pi^+$ for $\Omega_c^0\to\Xi^-K^-\pi^+\pi^+$ decays, and (d) $K^-\pi^+$ for $\Omega_c^0\to\Xi^0K^-\pi^+$ decays. 
The blue dotted lines show the signals, the green dashed lines show the background, and the solid lines the sum of the two.
Data are shown with circles. }
\label{fig:FigureRes}
\end{figure*}

Many of the modes under consideration may have resonant substructure that can help reveal their decay
mechanisms. Figure~\ref{fig:FigureRes}(a) shows 
the $\pi^+\pi^0$ invariant mass for the combinations within 
22 ${\rm MeV}/c^2$ 
($\approx 90\%$ efficient) of the $\Omega_c^0$ peak in the $\Omega_c^0\to\Omega^-\pi^+\pi^0$ mass distribution.
This distribution has been background-subtracted using  events from scaled sidebands between 
32  and $76\ {\rm MeV}/c^2$ from the peak. A fit is made to this distribution using
the sum of a $\rho^+$ signal shape and a nonresonant shape flat in phase space. The very small
efficiency difference between these two distributions is taken into account to calculate that $(83\pm10\%)$ of the
$\Omega^-\pi^+\pi^0$ mode proceeds via the $\rho^+$.
This result is consistent with the saturation of the $\Omega\pi^+\pi^0$ decay by the  
pseudo-two-body $\Omega^-\rho^+$ channel. We calculate a lower limit for the $\Omega^-\rho^+$ fraction by
integrating the likelihood function obtained from the fit, 
and finding the value of the fraction for which the integral contains 90\% of the 
total area.
This 90\% 
confidence-level lower limit value on the $\Omega^-\rho^+$ fraction of $\Omega^-\pi^+\pi^0$ is 71\%.

For the mode $\Omega_c^0\to\Xi^-K^-\pi^+\pi^+$, we define signal candidates as those within 7 ${\rm MeV}/c^2$ of the $\Omega_c^0$ mass;  
sidebands of $12-26\ {\rm MeV}/c^2$ from the $\Omega_c^0$ peak value; and present the 
scaled sideband-subtracted $\Xi^-\pi^+$ and $K^-\pi^+$ 
invariant mass distributions in Figs.~\ref{fig:FigureRes}(b) and \ref{fig:FigureRes}(c). Each distribution has 
two entries per $\Omega_c^0$ candidate. 
Polynomial nonresonant functions are fit to these distributions 
to find the yield of $\Xi^0(1530)$ and $\bar{K}^{*0}(892)$, respectively. Clear signals of $74\pm20$ events and $136\pm39$ events are found, 
where these uncertainties are statistical. 
These correspond to $(33\pm9)\%$ and  $(55\pm16)\%$ of the $\Xi^-K^-\pi+\pi^+$ decays proceeding through $\Xi^0(1530)$ and
$\bar{K}^{*0}(892)$, respectively.
There are indications that the signals include pseudo-two-body decays of the type 
$\Omega_c^0\to\Xi^{0}(1530)\bar{K}^{*0}(892)$, but the signal-to-noise ratio is not sufficient to allow for the measurement of
this process. Interference effects are expected to be small and are not taken into consideration.

For the mode $\Omega_c^0\to\Xi^0K^-\pi^+$, we select signal events within $11\ {\rm MeV}/c^2$ of the $\Omega_c^0$ peak value, 
and use sidebands of 22 to $44\ {\rm MeV}/c^2$. 
We then plot the sideband-subtracted $K^-\pi^+$ invariant mass distribution and observe a clear peak
due to the $\bar{K}^{*0}(892)$ meson. 
The sum of a $\bar{K}^{*0}(892)$ signal shape and a polynomial nonresonant shape is fit to this distribution and 
shown in Fig.~\ref{fig:FigureRes}(d).
The signal yield is determined to be $95 \pm 16$ events, corresponding to $(57\pm 10)\%$ of $\Xi^0K^{-}\pi^+$ decays. 

\section*{Systematic Uncertainties}

The systematic uncertainties that enter this analysis of the branching fractions are 
summarized in Table~\ref{tab:systematics}.
 To estimate the uncertainty due to the choice
of background shape, the order of the Chebyshev polynomial is increased by one and the change in yield taken as the systematic
uncertainty. As this always reduces the yield, this
is not done for the $\Omega_c^0\to\Sigma^+K^-K^-\pi^+$ mode, for which only an upper limit is quoted.
The sensitivity to the signal shape is found by repeating the analysis
with single, rather than double, Gaussian signal functions both for the normalizing mode and the signal mode. The MC simulation
program is tested using many similar reconstructed signals, and in all cases the extracted resolution values agree with the data within 10\%.
The systematic uncertainty
due to uncertainties in the resolution width are estimated from the change in yield when adjusting the signal widths by 10\%.

In addition, there are uncertainties in the simulation of the reconstruction efficiency 
that are not specific to this analysis. Care is taken to account for the cancelation of uncertainties in the calculation of
the branching ratios with respect to the normalizing mode.
We assign a relative uncertainty on the track reconstruction 
varying from 0.35\% 
to 2.5\%~\cite{CHANG1}. The relative uncertainties on the $\Lambda$, $K^0_S$, and $\pi^0$ reconstruction are 
4.0\%~\cite{CHANG1}, 2.8\%~\cite{SHEN}, 
and 3\%~\cite{CHANG2}, respectively. We use studies of $\Lambda\to p\pi^-$ and $D^0\to K^-\pi^+$ decays to assign uncertainties on the 
PID identification of the kaons and protons of 1.3\% per track~\cite{CHANG1}.   

Lastly, there is an uncertainty due to changes in the efficiencies when resonant substructure is present. As visible resonant substructure
is already taken into account in the efficiency calculations, this effect is small.
In the determination of the fractions due to substructure, the statistical uncertainties dominate over the small 
systematic uncertainties. 
The small differences in the efficiencies between the resonant and multi-body decays are taken into account in calculating the resonant
contribution to these modes.

\begin{table*}[htb]
\caption{The summary of the relative uncertainties (in \%). The systematic uncertainties are 
added in quadrature to give the last column.}
\begin{tabular}
 {     c          |c             |c       |c      |c          |c      |c     |c        |c          |c           |c}
\hline \hline
Mode               & { Statistical} & Bkgd &Signal & Signal   & Track & $K^0_S/\Lambda$& PID & $\pi^0$ & Resonances & {Total} \\
                   & { uncertainty} & shape &shape  & width   & finding &finding & requirements      & finding &                       &{ systematic}\\ 
\hline
$\Omega^-\pi^+\pi^0$      & {\bf 8.7 }   & 0.6  &0.3  & 4.2     & 0.0 & -    & -    & 3.0  &   1.0 & {\bf 5.3 } \\
$\Omega^-\pi^+\pi^-\pi^+$ & {\bf 15.0 }  & 2.3  &2.0  & 5.0     & 0.7 & -    & -    & -    &  3.0  &  {\bf  6.6      }  \\
$\Xi^-K^-\pi^+\pi^+$      & {\bf 10.6 }  & 0.6  &0.3  & 4.8     & 0.7 & -    & -    & -    &  1.0  &  {\bf   5.0     }  \\
$\Xi^0K^-\pi^+$           & {\bf 13.1 }  & 2.9  &0.5  & 4.2     & 2.5 & -    & -   & 3.0  &  2.0  &    {\bf 6.7    }  \\
$\Xi^-\bar{K^0}\pi^+$     & {\bf 11.1 }  & 3.4  &0.3  & 4.9     & 0.7 & 2.8  & 1.3   & -    & 1.0   &   {\bf 6.8    }   \\
$\Xi^0\bar{K^0}$           & {\bf 15.7 }  & 2.2 & 1.9  & 4.7    & 2.5 & 2.8  & 1.3   & 3.0  &  -    &  {\bf 7.4   }    \\
$\Lambda\bar{K^0}\bar{K^0}$ &{\bf 19.3 } &1.1  & 0.4  & 4.7     & 3.1 & 5.6  & 1.3   & -    & -     &   {\bf 8.1  }    \\
$\Sigma^+K^-K^-\pi^+$     & {\bf 50.9 }  & -   & 10.7  & 2.9 & 5.0     & 4.0  & 2.6  & 3.0  &  3.0  &  {\bf  13.6  }   \\
\hline 

\hline
\end{tabular}
\label{tab:systematics}
\end{table*}

\section*{Final Results}

The results for the branching fractions are summarized in Table~\ref{tab:answers}. 
In the case of $\Omega_c\to\Sigma^+K^-K^-\pi^+$, there is no significant signal. We calculate a 
90\% confidence upper limit by first combining the statistical and systematic uncertainties, 
and integrating the resultant likelihood function starting at $N_{\rm signal} = 0$; the upper limit is set
when the integral reaches 90\% of the total area.
For the cases where substructure
is measured, the 
fraction of the primary mode is given. 
The results assume a branching 
fraction $\bar{K^0}\to K^0_S$ of 50\%.

Four of the modes presented here have been measured previously~\cite{CLEO,BaBar,E687}. In all cases, 
these new measurements are consistent, within two standard deviations, 
with the previous measurements~\cite{PDG} and provide substantial improvements 
in precision.
It is surprising that we find a restrictive limit on the decay $B(\Omega_c\to\Sigma^+K^-K^-\pi^+)/B(\Omega^-\pi^+$), 
even though the E687 experiment, albeit with different relative efficiencies, finds a much larger signal in $\Sigma^+K^-K^-\pi^+$
than $\Omega^-\pi^+$.

There is a paucity of recent predictions on the branching fractions of charmed baryons. 
However, some patterns in the data of charmed baryon decays are clear. Whereas the other
weakly decaying charmed baryons $Y_c$ have branching ratios 
${\cal B}(Y_c \to Y\pi^+\pi^-\pi^+)/{\cal B}(Y_c \to Y\pi^+) \gg 1$, it is confirmed that, when 
$Y_c$ is an $\Omega_c$, this ratio is considerably less than 1. While multi-body weak decays are difficult to model
theoretically, we hope that these new results on pseudo-two-body decays will spur further theoretical work.

\begin{table*}[htb]
\caption{The summary of the results to the fits shown in Figs.~\ref{fig:Figure1}, \ref{fig:Figure2}, 
and~\ref{fig:FigureRes}. The numbers
in parentheses refer to the fraction of the multi-body final state that includes the listed resonance.} 

\begin{tabular}
 {     c          |c             |c       |c          }

\hline \hline
Mode                     & Branching ratio & Substructure & Previous measurement  \\
                         & with respect to $\Omega^-\pi^+$ &                      \\
\hline
$\Omega^-\pi^+     $      & 1                 &   &                                      \\
$\Omega^-\pi^+\pi^0$      & $2.00\pm0.17\pm0.11 $   &   & $1.27\pm0.3\pm0.11$~\cite{BaBar}  \\
$\Omega^-\rho^+$          &   & $>71\%$ &      \\
$\Omega^-\pi^+\pi^-\pi^+$ & $0.32\pm0.05\pm0.02 $   &  & $0.28\pm0.09\pm0.01$~\cite{BaBar}  \\
$\Xi^-K^-\pi^+\pi^+$      & $0.68\pm0.07\pm0.03$  &   & $0.46\pm0.13\pm0.03$~\cite{BaBar}  \\
$\Xi^{0}(1530) K^-\pi^+$ &  & $ (33\pm9)\%$     &   \\
$\Xi^-\bar{K}^{*0}\pi^+$        & & $ (55\pm16)\%$    &   \\
$\Xi^0K^-\pi^+$           & $1.20\pm0.16\pm0.08$  &   & $4.0\pm2.5\pm0.4$~\cite{CLEO}  \\
$\Xi^{0}{\bar{K}^{*0}}$           &   &$(57\pm 10)\%$   &   \\
$\Xi^-\bar{K^0}\pi^+$       & $2.12\pm0.24\pm0.14$  &   &   \\
$\Xi^0\bar{K^0}$            &$1.64\pm0.26\pm0.12$              &   & \\
$\Lambda\bar{K}^0\bar{K}^0$ &$1.72\pm0.32\pm0.14$ &   &   \\
$\Sigma^+K^-K^-\pi^+$     &  $ <0.32 $ (90\% CL)  &  &    \\

\hline
\hline

\end{tabular}

\label{tab:answers}
\end{table*}

\section*{Acknowledgments}

We thank the KEKB group for the excellent operation of the
accelerator; the KEK cryogenics group for the efficient
operation of the solenoid; and the KEK computer group,
the National Institute of Informatics, and the 
PNNL/EMSL computing group for valuable computing
and SINET5 network support.  We acknowledge support from
the Ministry of Education, Culture, Sports, Science, and
Technology (MEXT) of Japan, the Japan Society for the 
Promotion of Science (JSPS), and the Tau-Lepton Physics 
Research Center of Nagoya University; 
the Australian Research Council;
Austrian Science Fund under Grant No.~P 26794-N20;
the National Natural Science Foundation of China under Contracts 
No.~10575109, No.~10775142, No.~10875115, No.~11175187, No.~11475187, 
No.~11521505 and No.~11575017;
the Chinese Academy of Science Center for Excellence in Particle Physics; 
the Ministry of Education, Youth and Sports of the Czech
Republic under Contract No.~LTT17020;
the Carl Zeiss Foundation, the Deutsche Forschungsgemeinschaft, the
Excellence Cluster Universe, and the VolkswagenStiftung;
the Department of Science and Technology of India; 
the Istituto Nazionale di Fisica Nucleare of Italy; 
National Research Foundation (NRF) of Korea Grants No.~2014R1A2A2A01005286, No.~2015R1A2A2A01003280,
No.~2015H1A2A1033649, No.~2016R1D1A1B01010135, No.~2016K1A3A7A09005603, No.~2016R1D1A1B02012900; Radiation Science Research Institute, Foreign Large-size Research Facility Application Supporting project and the Global Science Experimental Data Hub Center of the Korea Institute of Science and Technology Information;
the Polish Ministry of Science and Higher Education and 
the National Science Center;
the Ministry of Education and Science of the Russian Federation and
the Russian Foundation for Basic Research;
the Slovenian Research Agency;
Ikerbasque, Basque Foundation for Science and
MINECO (Juan de la Cierva), Spain;
the Swiss National Science Foundation; 
the Ministry of Education and the Ministry of Science and Technology of Taiwan;
and the U.S.\ Department of Energy and the National Science Foundation.


\begin{thebibliography}{99}
\bibitem{CC}
Throughout this paper, 
the inclusion of the charge-conjugate mode decay is implied
unless otherwise stated.
\end{thebibliography}
\end{document}